# Giant Convection Cells Found on the Sun

David H. Hathaway[1]*, Lisa Upton[2,3] and  Owen Colegrove[4]

[1]NASA Marshall Space Flight Center, Huntsville, AL 35812 USA. [2]Vanderbilt University, Nashville, TN 37235 USA. [3]The University of Alabama in Huntsville, Huntsville, AL 35899 USA. [4]University of Rochester, Rochester, NY 14627 USA.

*To whom correspondence should be addressed. E-mail: david.hathaway@nasa.gov

Abstract:  Heat is transported through the outermost 30% of the Sun's interior by overturning convective motions. These motions are evident at the Sun's surface in the form of two characteristic cellular structures – granules and supergranules (~1000 and ~30,000 km across respectively). The existence of much larger cells has been suggested by both theory and observation for over 45 years. We found evidence for giant cellular flows that persist for months by tracking the motions of supergranules. As expected from the effects of the Sun's rotation, the flows in these cells are clockwise around high pressure in the north, counter-clockwise in the south and transport angular momentum toward the equator, maintaining the Sun's rapid equatorial rotation.

The Sun, like most stars, has an outer convection zone in which heat generated by nuclear reactions in its core is transported to its surface by overturning convective motions. These motions were evident in early telescopic observations of the Sun as granules, which are bright grain-like structures with typical diameters of ~1000 km, lifetimes of ~10 min, and flow velocities of ~3000 m s$^{-1}$. Much larger structures – supergranules - were evident from their flow velocities as seen in the Doppler shift of atomic spectral lines formed in the Sun's surface layers (1, 2). Supergranules have diameters ~30,000 km, lifetimes of ~24 hours, and flow velocities of ~500 m s$^{-1}$. Both granules and supergranules cover the entire solar surface but are substantially modified by the intense magnetic fields in and around sunspots.

The existence of even larger convection cells - giant cells - was proposed shortly after supergranules were detected (3). These cells are expected to span the 200,000-km-deep solar convection zone, to have diameters of ~200,000 km and lifetimes of ~1 month, and to be heavily influenced by the Sun's 27-day rotation. Hydrodynamical models of convective motions in the Sun's rotating convection zone (4-6) suggest that these cells should be elongated north-to-south near the equator and be sheared off at higher latitudes by the Sun's differential rotation (the equatorial regions rotate once in ~25 days while the polar regions rotate once in ~35 days). These "banana" cells should transport angular momentum toward the Sun's equator - a critically important process for maintaining the differential rotation.

The observational evidence for the existence of giant cells has been only suggestive. Magnetic structures of a similar size and shape have been observed (7) but these structures are fully explained (8) by the transport of magnetic elements away from active region sunspots by well-characterized flows: differential rotation, supergranules, and the poleward meridional circulation. The best evidence for the existence of giant cells are observations indicating that large-scale velocity features do exist in the spectrum of motions and are moving with the Sun's rotation (9-11).

We measure the motions of the supergranules themselves, with the expectation that the supergranules will be carried from the centers to the boundaries of the giant cells by these larger, long-lived flows. We obtained images of the line-of-sight Doppler shifts of a spectral line formed by traces of iron in the Sun's lower atmosphere with the Helioseismic and Magnetic Imager (HMI) on the NASA Solar Dynamics Observatory (SDO) every 45 s (12). These 4096- by 4096-pixel images were averaged over 12 min, blurred over 11-by-11 pixels, and resampled at 512- by 512-pixel resolution for our measurements. The line-of-sight motion of the HMI instrument relative to the Sun was removed as was the Doppler signal due to the Sun's solid-body rotation and imaging artifacts produced in the instrument itself. The data were then mapped to heliographic longitude and latitude. Two large-scale Doppler velocity signals (an east-west gradient due to the Sun's differential rotation relative to the solid-body rotation and a disk-center-to-limb variation due to the correlation between radial up-flow and brightness in the granules) were then measured and removed from the data so as to fully isolate the supergranule flow structures (Fig. 1).

We generated these images of supergranules hourly starting in May 2010. We determined the motions of the supergranules by local cross-correlation tracking (13) using image pairs separated by 8, 16, and 24 hours. We cross-correlated the signal in 21- by 21-pixel blocks in the earlier image with similar blocks in the later image to find the displacement that gives the highest correlation. We set a lower limit to acceptable correlations. This had the effect of eliminating the most uncertain measurements - primarily from disk center where the Doppler signal due to these horizontal flows is weak. We determined the displacements to within a fraction of a pixel using a parabolic fit to the correlations about the peak. These displacements yield velocities in longitude and latitude for the group of supergranules covered by the pixel block at each location. This process yields hourly 256- by 256-pixel images of the flow velocities of the supergranules. We averaged these flow velocity images over each 27-day solar rotation using the longitude of the central meridian to position them relative to the other hourly velocity images. This typically gave an average over ~300 hours at each location. These supergranule flow velocity maps are dominated by the axisymmetric flows: differential rotation and meridional flow. Removing these longitudinally averaged velocities reveals the giant cells as large-scale and long-lived velocity structures in the supergranule flow velocity maps (Fig. 2 and figs. S1 and S2).

The most striking features are those seen at higher latitudes. The longitudinal velocity maps show velocity structures which are swept back in longitude at higher latitudes in each hemisphere. These features persist and drift in longitude by ~180° over the three 27-day rotation intervals, indicating lifetimes of at least three months and a rotation period of ~32 days at those latitudes. The structures at lower latitudes are less well defined and shorter lived. However, although less pronounced, the low latitude structures seen in the latitudinal velocity maps do appear to be aligned north-to-south. A low latitude north-south alignment has also been noted in the structure of the supergranulation pattern itself (14).

We find virtually the same cellular patterns for all three time lags (fig. S3) but with weaker flow velocities from the shorter time lags. The root-mean-square velocities are 16 m s$^{-1}$ with 24-hour time lags but only 10 m s$^{-1}$ and 8 m s$^{-1}$ at 16-hour and 8-hour time lags respectively. The measured differential rotation and meridional flow also vary systematically with increasing time lag (rotation rate increases while the meridional flow velocity decreases). These variations are thought to be due to flow variations with depth (15, 16). The correlations at longer time lags are dominated by larger supergranules that live longer, extend deeper into the Sun, and are transported by the flows at those greater depths (~50 Mm for the 24-hour time lag). This implies that the giant cell flow velocities decrease in amplitude as they approach the surface and it helps to explain the low upper limits on giant cell flow velocities given by previous searches (17, 18). Measurements with even longer time lags are possible, but the correlations are much weaker and give noisier results.

The preferential visibility of east-west structures in the longitudinal flow and north-south structures in the latitudinal flow is one indication of the effects of the Sun's rotation on these large-scale flows. The Coriolis force due to the Sun's rotation turns the flow velocities to be more parallel to these elongated structures. In addition, we found that the kinetic helicity – the correlation between diverging flows and flow vorticity - is negative in the north and positive in the south (Fig. 3). A more meaningful indication of the effect of the Sun's rotation is seen in the Reynolds stress component, $\langle V_\phi V_\theta \rangle$, the correlation between longitudinal and latitudinal flows (Fig. 4). This stress is necessary in the hydrodynamical models (4-6) in order to produce a rapidly rotating equator. The presence of these statistical correlations in the observed flows is further evidence that we have indeed found giant convection cells.

While many of the large-scale magnetic structures initially attributed to giant cells (6) can be explained with magnetic flux transport by other well characterized flows (7), the initial formation of active regions may nonetheless be associated with these giant cell flows. On one hand, active region formation may be favored in diverging flows in which upflows carry the magnetic field to the surface. On the other hand, active regions may form in converging flows in which magnetic fields become concentrated. Our initial search for correlations between converging/diverging flows and active regions formation has been inconclusive. However, it would be surprising if these large-scale, long-lived flows did not substantially influence the evolution and structure of the Sun's magnetic field.

The SDO/HMI data described in this paper are archived at: http://jsoc.stanford.edu/. The SDO/HMI project is supported by NASA grant to Stanford University. D. Hathaway was supported by a grant from the NASA Heliophysics Supporting Research and Technology (SR&T) Program to NASA/MSFC. L. Upton was supported by a grant from the NASA Living With a Star (LWS) Program to NASA/MSFC. O. Colegrove was supported as a Research Experience for Undergraduates (REU) summer student at the University of Alabama in Huntsville by funds from NSF Grant No. AGS-1157027.

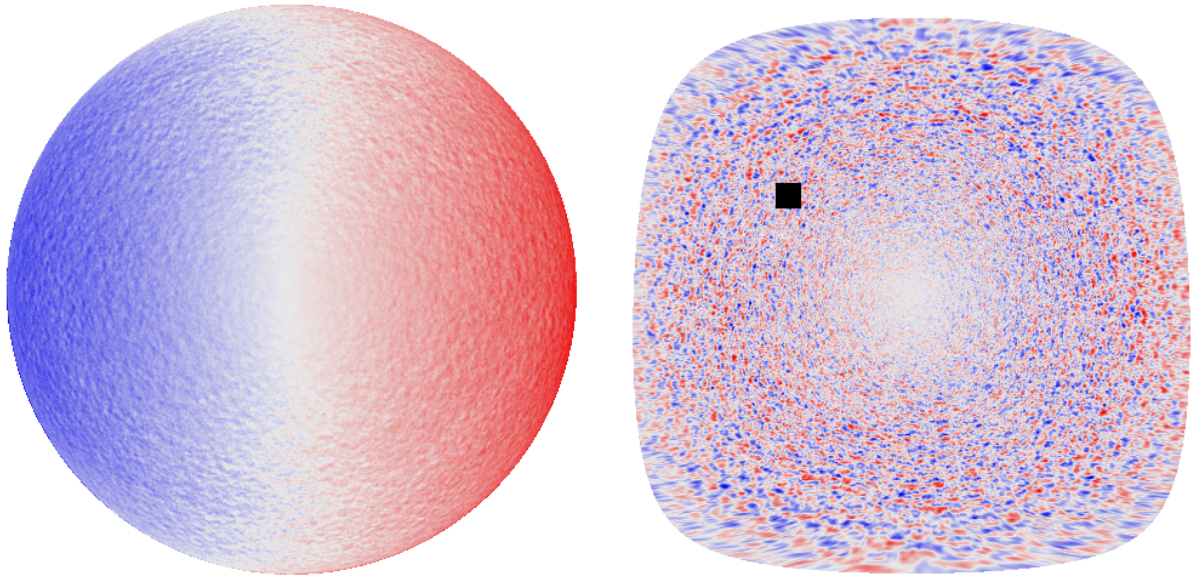

**Fig. 1. Doppler velocity images. (Left)** a 12-minute average Doppler velocity image from the HMI instrument with red representing red-shifted pixels and blue representing blue-shifted pixels (with a velocity range of ±3000 m s$^{-1}$). **(Right)** the same data mapped to heliographic longitude and latitude with the instrumental signals and global flows removed to isolate the pattern of supergranule cells (with a velocity range of ±600 m s$^{-1}$). The black square shows the size of the block of pixels used in the local correlation tracking procedure.

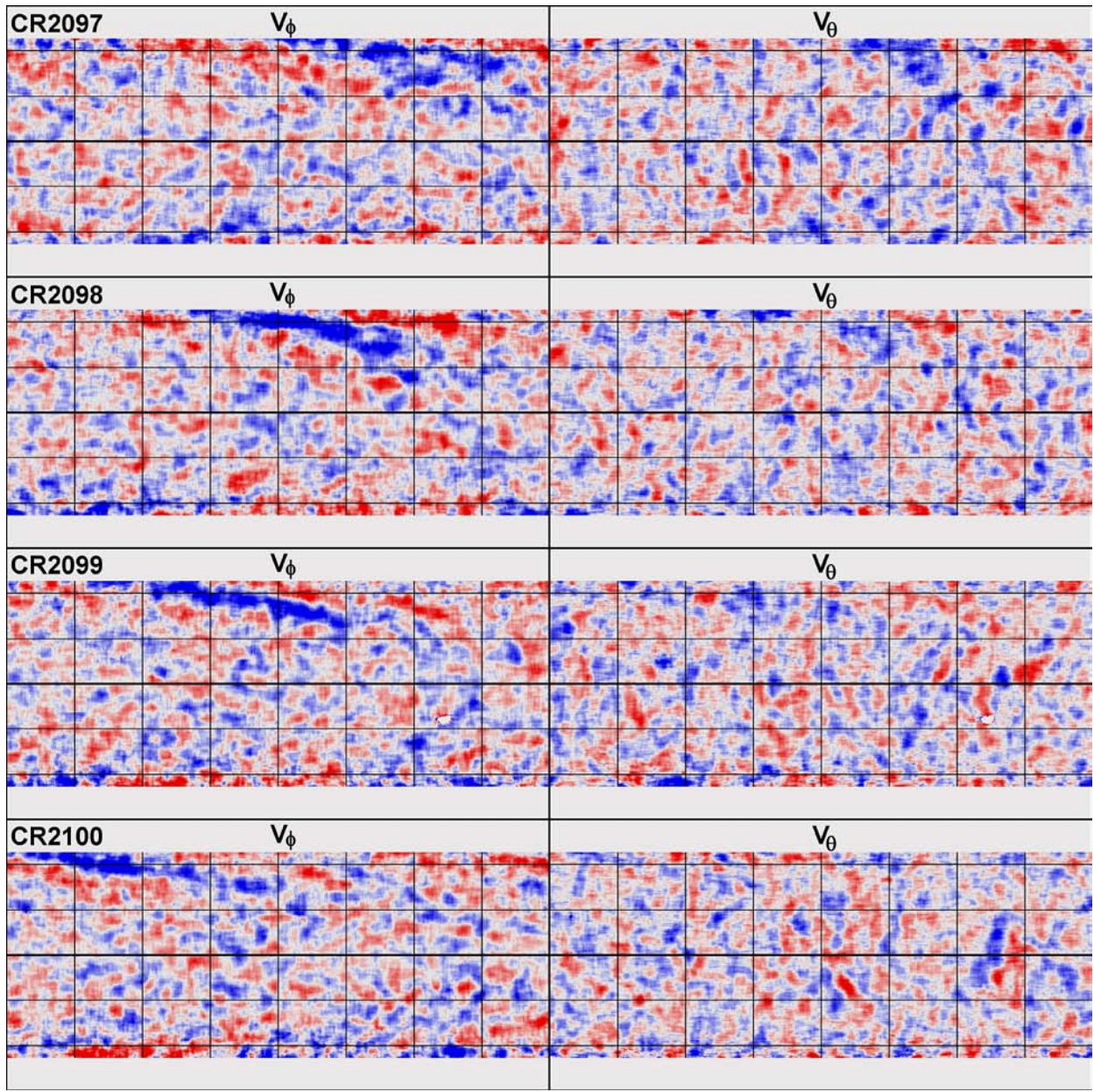

**Fig. 2. Supergranule flow velocity maps.** This sequence (top-to-bottom) of Mercator projection maps of the longitudinal **(left)** and latitudinal **(right)** velocity of the supergranules was obtained from four rotations of the Sun from May to August 2010 (prograde and southward velocities are red and retrograde and northward velocities are blue with a range of ±20 m s$^{-1}$). All maps cover the full 360° of longitude but are limited to ±70° latitude. The vertical lines are at 45° longitude intervals whereas the horizontal lines are at 30° latitude intervals.

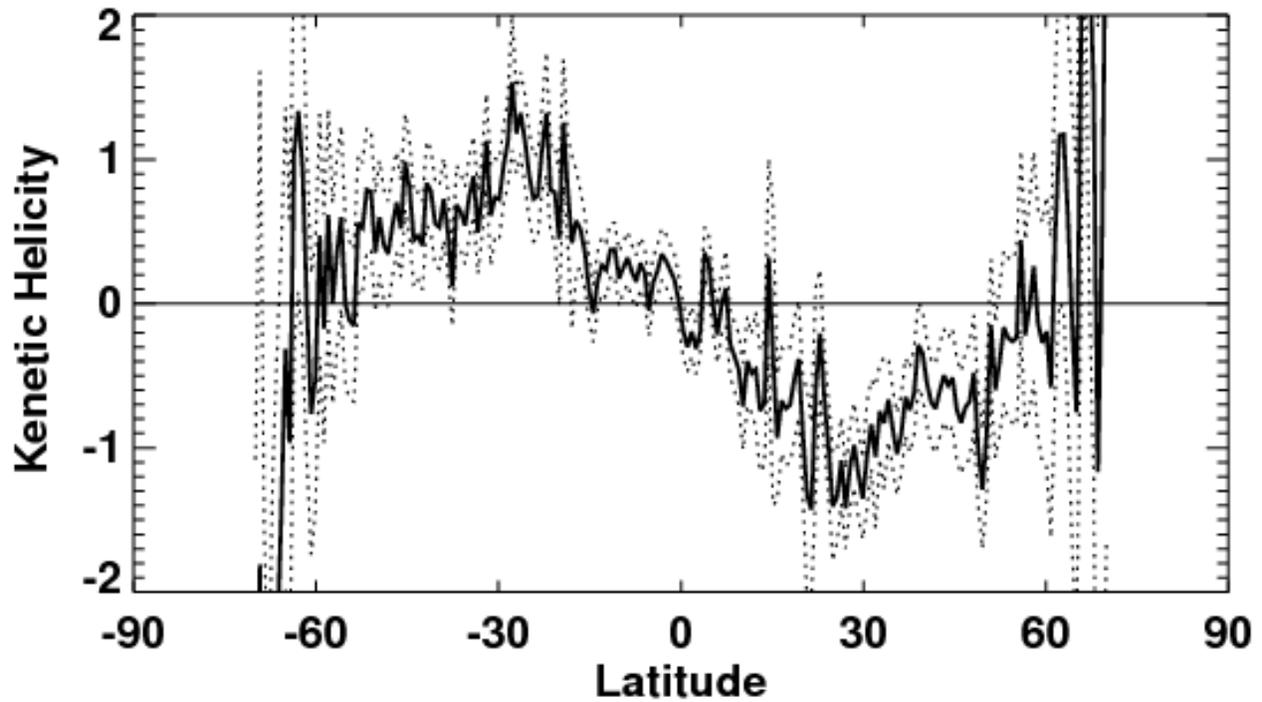

Fig. 3. The kinetic helicity produced by the giant cells as a function of latitude with 2σ error limits from the first two years of HMI. Kinetic helicity in arbitrary units; solid lines indicate the signal, and dotted lines indicate 2σ error limits. The correlation between diverging flows and vertical vorticity is negative in the northern hemisphere and positive in the southern hemisphere. This indicates clockwise circulation around high pressure centers of divergence in the north and counter-clockwise circulation around those in the south.

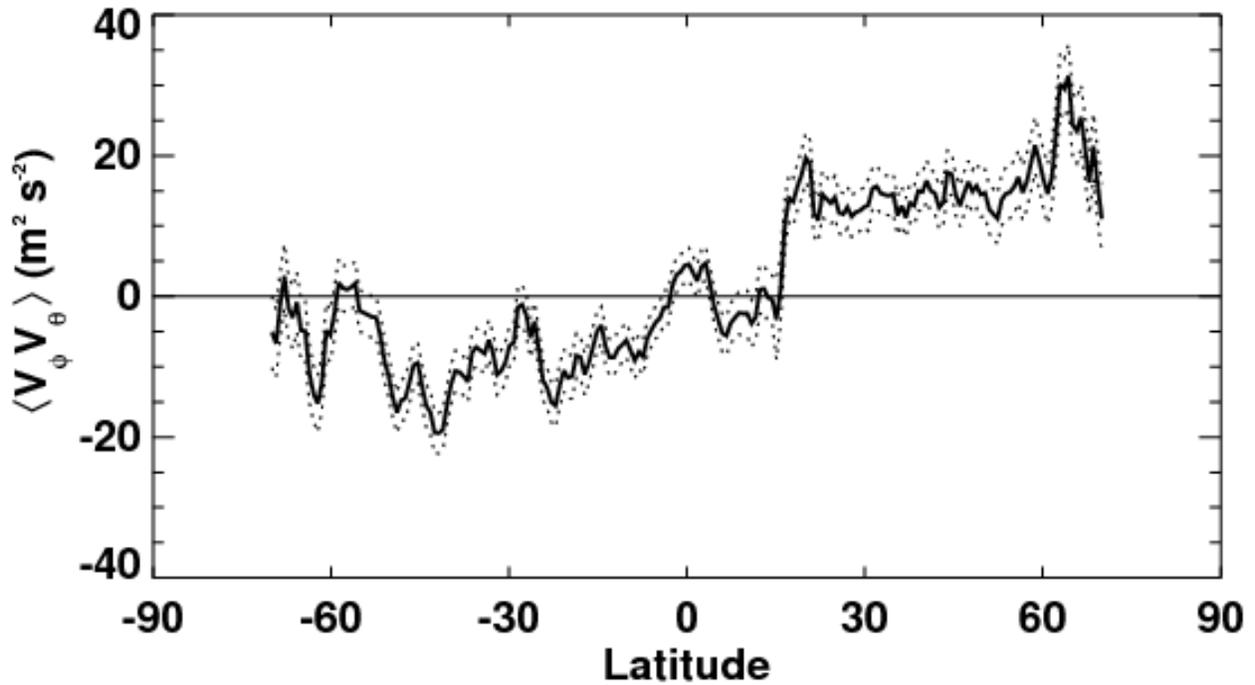

**Fig. 4. The Reynolds stress, $\langle V_\phi V_\theta \rangle$, produced by the giant cells as a function of latitude with 2σ error limits from the first two years of HMI.** Solid lines indicate the Reynolds stress signal, and dotted lines indicate 2σ error limits. The correlation between prograde flow (positive $V_\phi$) and southward flow (positive $V_\theta$) is positive in the northern hemisphere and negative in the southern hemisphere. This indicates a transport of angular momentum equatorward, which is the required direction for maintaining the more rapid rotation of the Sun's equator.

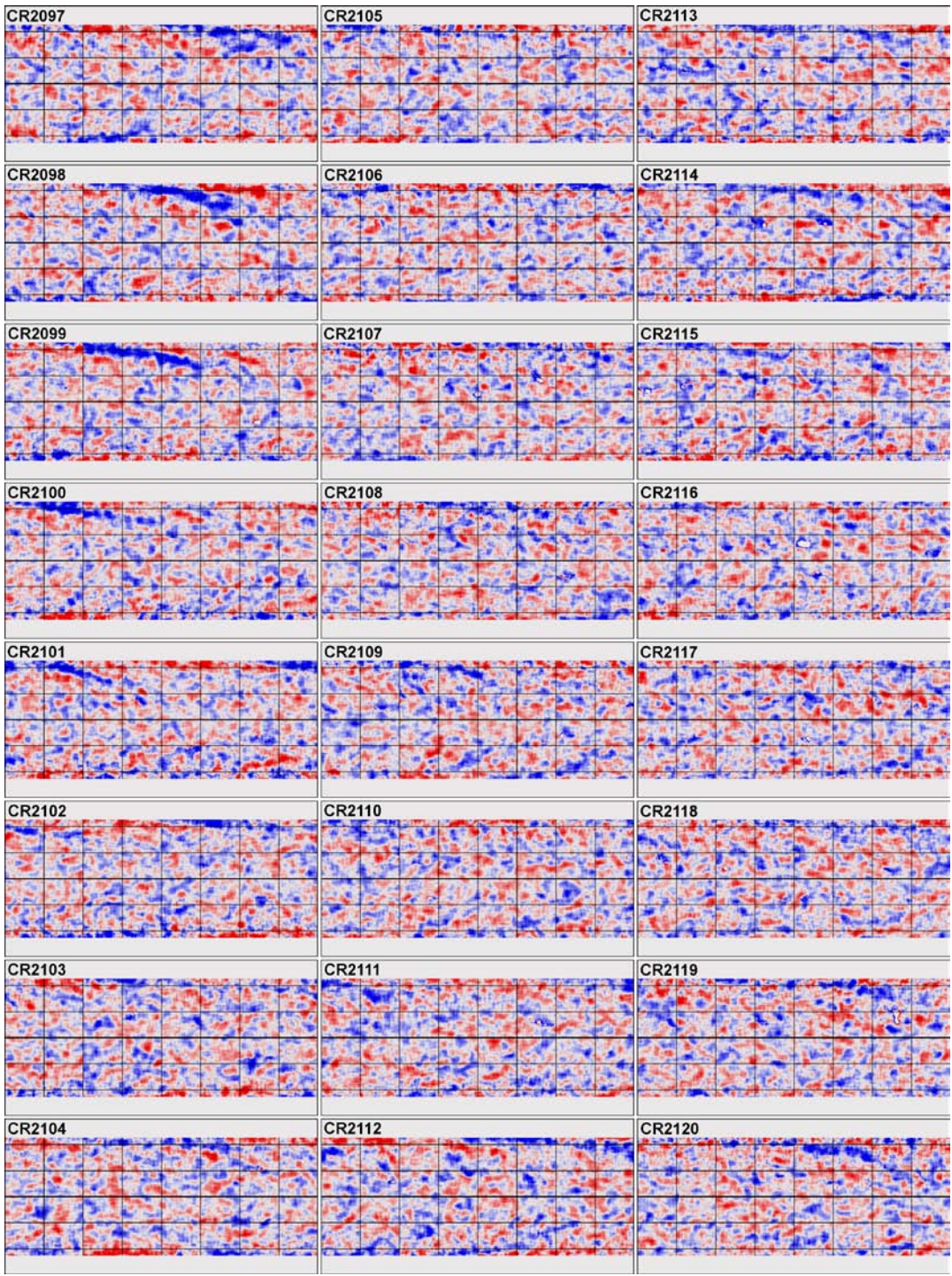

**Fig. S1**
Supergranule longitudinal flow velocity maps for the first 24 solar rotations (May 2010 to March2012) observed with HMI. Many features persist and drift with the Sun's differential rotation for several months.

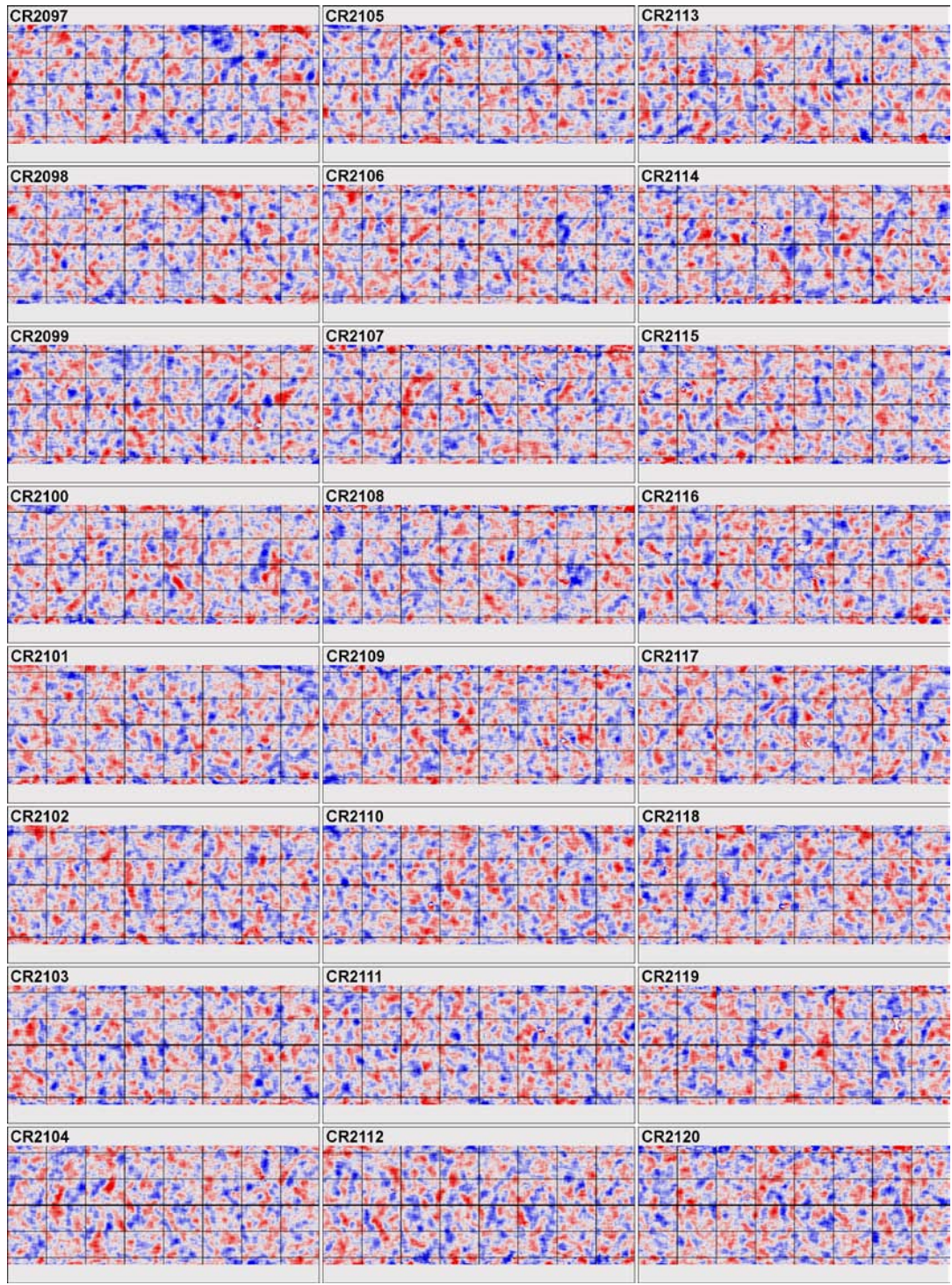

**Fig. S2**
Supergranule latitudinal flow velocity maps for the first 24 solar rotations (May 2010 to March2012) observed with HMI. Many features persist and drift with the Sun's differential rotation for several months.

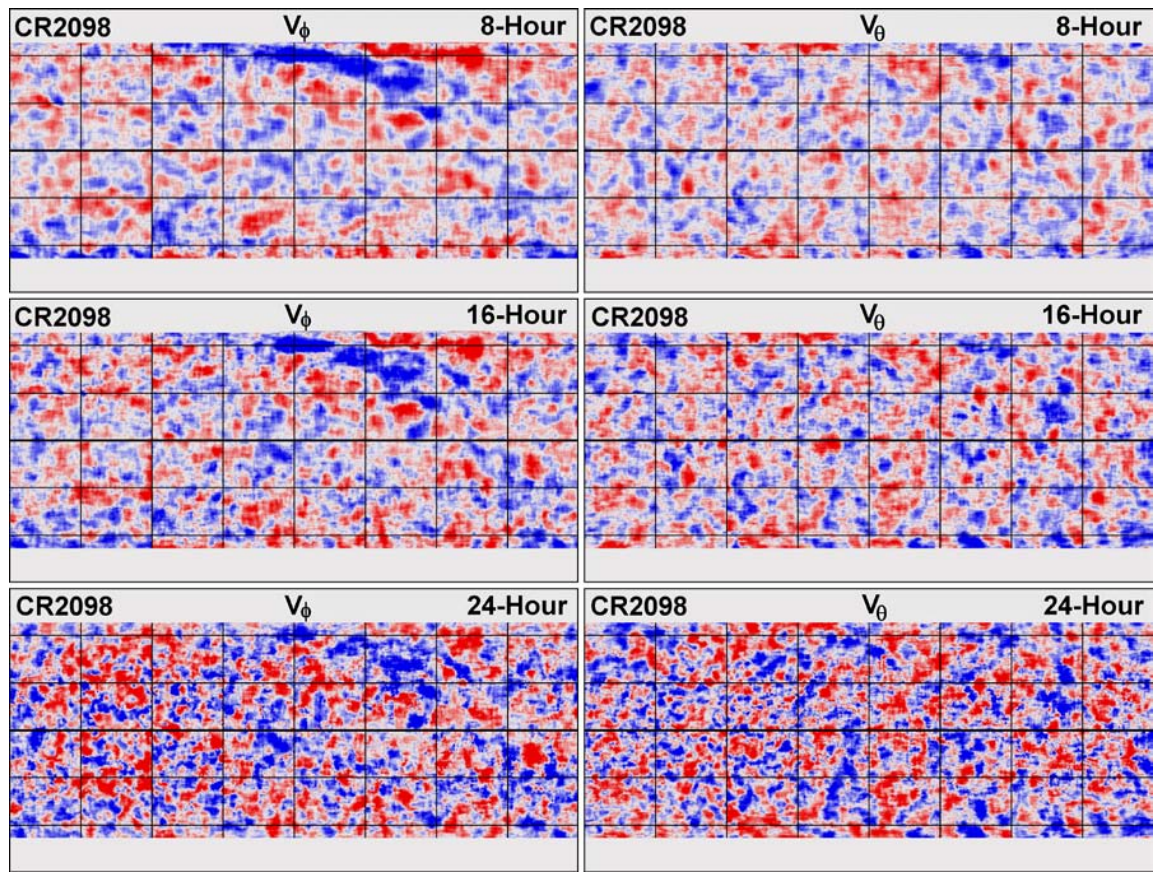

**Fig. S3.**

Supergranule flow velocity maps for different time-lags. The longitudinal (left) and latitudinal (right) velocity of the supergranules were measured using three different time-lags for the cross-correlation: 8 hours (top), 16 hours (middle), and 24 hours (bottom). The same velocity pattern is obtained with all three time lags. The primary difference is an increase in velocity amplitude with time-lag.